\documentclass[aps,pra,
 twocolumn
  ,showpacs,superscriptaddress]{revtex4}
\usepackage{graphicx}
\usepackage{amsmath}
\usepackage{amssymb}
\usepackage{amsfonts}
\usepackage{natbib}
\usepackage[ps2pdf,breaklinks,colorlinks,
linkcolor=blue,citecolor=blue,urlcolor=blue]{hyperref}
\usepackage{dcolumn}
\newcolumntype{d}[1]{D{.}{.}{#1}}

\begin{document}
\title{
Energy and structural properties of $N$-boson clusters attached to 
three-body Efimov states: 
Two-body zero-range interactions and the role of the three-body regulator
}
\author{Yangqian Yan}
\affiliation{Department of Physics and Astronomy,
Washington State University,
  Pullman, Washington 99164-2814, USA}
\author{D. Blume}
\affiliation{Department of Physics and Astronomy,
Washington State University,
  Pullman, Washington 99164-2814, USA}
\date{\today}
\begin{abstract}
The low-energy spectrum of $N$-boson clusters with pairwise
zero-range interactions is believed to be governed by a three-body parameter.
We study the ground state of $N$-boson clusters with infinite two-body
$s$-wave 
scattering length by performing {\em{ab initio}} Monte Carlo
simulations.
To prevent Thomas collapse, different finite-range three-body regulators are
used.
The energy and structural properties for the three-body 
Hamiltonian with two-body
zero-range interactions and three-body regulator
are in much 
better agreement with the ``ideal zero-range Efimov theory'' results than
those for Hamiltonian with two-body finite-range interactions.
For larger clusters we find 
that the ground state energy and structural properties of the Hamiltonian with
two-body zero-range interactions and finite-range three-body regulators are not
universally determined by the three-body parameter, i.e., dependences on the
specific form of the three-body regulator are observed.
For comparison, we consider Hamiltonian with two-body van der Waals
interactions and no three-body regulator. For the interactions considered,
 the ground state
energy of the $N$-body clusters is---if scaled by the three-body
ground state energy---fairly universal, i.e., the dependence on the
short-range details of the two-body van der Waals potentials is small.
Our results are compared with the literature.
\end{abstract}
\maketitle
\section{Introduction}
The unitary regime, where the two-body $s$-wave scattering length is
infinitely large, can be reached in ultra cold dilute atomic gases
using Feshbach resonance techniques~\cite{eita10}.
Two-component Fermi gases were realized experimentally
and found to be
stable and universal even in the large
$s$-wave scattering length regime~\cite{jin99,jin04,ketterle04}, i.e., the 
properties of the system were found to be governed,
to a very good approximation, by the $s$-wave scattering length $a_s$ alone and
independent of the details
of the interaction potential~\cite{stringariFermireview,
blochreview,blumerev12}. Unitary Bose gases, in
contrast, are short-lived~\cite{salomon13,hadzibabic13,jin14}.
Their properties depend on the details of the interaction 
potential.
Typically, this dependence is encapsulated by a three-body
parameter~\cite{BraatenHammerReview}.

Efimov predicted that three identical bosons interacting through
two-body potentials with infinitely large $s$-wave scattering length $a_s$
and vanishing 
effective range support
an infinite number of three-body bound states~\cite{efimov70}.
The binding momenta $\kappa_3^{(n)}$ of the
trimers ($n$ labels the states) display a geometric
scaling, i.e.,
$\kappa_3^{(n)}/\kappa_3^{(n+1)}
\approx22.6944$~\cite{efimov70,BraatenHammerReview}.
If the binding momentum of one trimer is known, that of the other trimers is 
also known.
Importantly, the binding momenta themselves
cannot be determined solely from a theory
that is based on two-body zero-range potentials. 
Rather, a three-body parameter is needed to regularize the problem (i.e., to
set the absolute scale of the three-body spectrum). The
three-body regulator can be introduced in many ways. In this work, we consider
three different regularization approaches: 
(i) a Hamiltonian with two-body zero-range potentials and a 
zero-range three-body potential,
(ii) a Hamiltonian with two-body zero-range potentials 
and a purely repulsive three-body potential,
and 
(iii) a Hamiltonian with finite-range 
two-body potentials and no three-body potential.

Much less is known about four- and higher-body systems at unitarity~\cite{
platter04,hanna06,hammer07,javier09,deltuva10,hadizadeh11,javier11,hadizadeh13,kievsky14}. 
$N$-body cluster states are believed to be attached to each trimer, i.e., for 
a trimer with binding momentum $\kappa_3^{(n)}$,
two $N$-body states are believed to exist with binding momenta $C_N^{(1)}
\kappa_3^{(n)}$ and $C_N^{(2)}\kappa_3^{(n)}$, where $C_N^{(1)}$ and
$C_N^{(2)}$ are dimensionless parameters 
that do not depend on $n$.
Whether four- and
higher-body parameters exist has been under debate in the
literature.

\begin{figure}[!tbp]
\centering
\includegraphics[angle=0,width=0.4\textwidth]{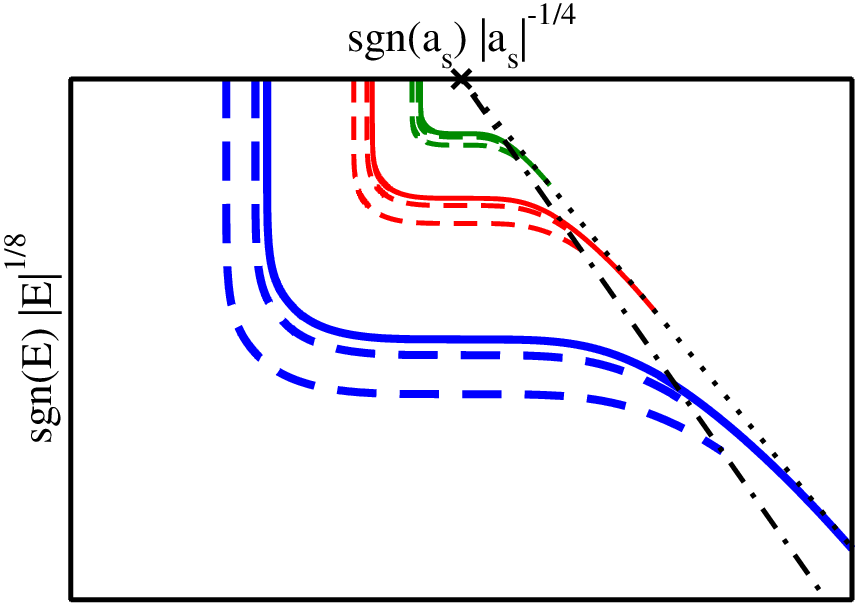}
\caption{(Color online)
  Schematic illustration of the energy spectrum for four identical bosons.
  The x marks the $(1/a_s,E)=(0,0)$ point.
  The dotted line shows the energy of the weakly-bound dimer. The solid lines
  show different Efimov trimer states, which become unbound on the positive
  scattering length side at the atom-dimer
  threshold. The dashed lines show ``ground state'' and ``excited state''
  tetramers that are attached to each Efimov trimer. These tetramer states hit
  the dimer-dimer threshold on the positive scattering length side (the energy
  of the two dimers is shown by the dash-dotted line). It should be noted that
  the excited tetramer state turns into a virtual state for a certain region
  of positive scattering lengths~\cite{deltuva11};
  this detail is not reflected in the plot.
 }\label{Fig1}
\end{figure}

The study of $N$-body states attached to Efimov trimers is challenging for
several reasons. To date, no analytical solutions for $N\ge4$ exist. Numerical
treatments have to be capable of describing vastly different length scales.
For finite-range two-body interactions, the lowest trimer state is typically
not a ``pure'' Efimov state. Thus, one would ideally like to investigate
$N$-body droplets that are tied to the first- or second-excited trimer states.
The corresponding $N$-body states ($N\ge4$; see Fig.~\ref{Fig1}
for an
illustration of the four-body spectrum as a function of $1/a_s$) are not
bound states but resonance
states, which are not stable with respect to break-up into smaller sub-units.
Thus, the numerical approach of choice would ideally be capable of treating
$N$-body resonance states whose size is many orders of magnitude larger than
the range of the underlying two-body potential. 

To bypass these numerical
challenges, this work pursues, as have other works 
before~\cite{javier10,kievsky14}, an approach
that considers $N$-body droplets (the thick dashed lines in
Fig.~\ref{Fig1} show the two four-body states) tied to the energetically
lowest-lying trimer state (thick solid line in Fig.~\ref{Fig1}). To ensure
that the trimer ground state has the key
characteristics of a true Efimov trimer state, we employ two-body zero-range
interactions together with a purely repulsive three-body potential that serves
as a regulator; we refer to this model as 2bZR+3bRp (2b, ZR, 3b, and R stand
for two-body, zero-range, three-body, and repulsive, respectively, and $p$
denotes the power of the repulsive three-body potential; see below).
The forms of $V_{\text{2b}}$ and $V_{\text{3b}}$ for the model 2bZR+3bRp are given in
Table~\ref{tab1} and 
\begin{table}[!tbp]
  \centering
  \caption{
Summary of potential models considered in this work. For each model, the
two-body potential $V_{\text{2b}}$ and the three-body potential $V_{\text{3b}}$ are
listed. 
$V_{\text{2b}}$ for 2bZR+3bZR, 2bZR+3bHC, and 2bZR+3bRp is the Fermi-Huang
pseudopotential~\cite{yang57}; $a_s$ is set to infinity.
$V_{\text{ZR}}(R)$ for 2bZR+3bZR is treated as a zero-range boundary condition.
$V_{\text{HC},R_0}(R)$ is the hardcore repulsive potential; $V_{\text{HC},R_0}(R)=0$ for
$R>R_0$ and $V_{\text{HC},R_0}(R)=\infty$ for $R<R_0$.
$V_0$ and $r_0$ for  2bG, 
 $c_{12}$ and $c_{6}$ for 2bLJ, $c_{10}$ and $c_{6}$ for 2b10-6,
and $c_{8}$ and $c_{6}$ for 2b8-6 are chosen such that the $s$-wave
 scattering length is
infinitely large and the two-body system supports one zero-energy $s$-wave
bound state.
}
\begin{tabular} 
{c c c}
    \hline
    \hline
     model &$V_{\text{2b}}$ & $V_{\text{3b}}$        \\
    \hline
    2bZR+3bZR  & $\frac{4\pi\hbar^2}{m}a_s \delta^{(3)}
    (\mathbf{r})\frac{\partial}{\partial r}r$ & $V_{\text{ZR}}(R)$     \\ 
    2bZR+3bHC &$\frac{4\pi\hbar^2}{m}a_s
    \delta^{(3)}(\mathbf{r})\frac{\partial}{\partial r}r$ & $V_{\text{HC},R_0}(R)$ \\
    2bZR+3bRp& $\frac{4\pi\hbar^2}{m}a_s \delta^{(3)}(\mathbf{r})
    \frac{\partial}{\partial r}r$ & $\frac{C_p}{R^p}$        \\ 
    2bG &$V_0\exp[-r^2/(2r_0^2)]$ & ---     \\ 
    2bLJ & $\frac{c_{12}}{r^{12}}-\frac{c_6}{r^6}$ & ---     \\ 
    2b10-6 & $\frac{c_{10}}{r^{10}}-\frac{c_6}{r^6}$ & ---     \\ 
    2b8-6 & $\frac{c_{8}}{r^{8}}-\frac{c_6}{r^6}$ & ---     \\ 
    \hline
    \hline
  \end{tabular}
  \label{tab1}
\end{table}
the Hamiltonian $H$ for $N$ particles with mass $m$ and position vector
$\mathbf{r}_j$ reads
\begin{equation}
  H=
  -\sum_{j=1}^{N} \frac{\hbar^2}{2m}\nabla_j^2+
  \sum_{j<k}^{N}V_{\text{2b}}(\mathbf{r}_{jk})+\sum_{j<k<l}^{N}V_{\text{3b}}(R_{jkl}),
  \label{hamiltonian}
\end{equation}
where 
the two-body potential $V_{\text{2b}}$ depends on the interparticle distance vector
$\mathbf{r}_{jk}$ ($\mathbf{r}_{jk}=\mathbf{r}_j-\mathbf{r}_k$) and the
three-body potential $V_{\text{3b}}$ depends on the three-body hyperradius $R_{jkl}$,
\begin{equation}
R_{jkl}=\sqrt{(\mathbf{r}_{jk}^2+\mathbf{r}_{jl}^2+\mathbf{r}_{kl}^2)/3}.
\end{equation}
Importantly, the $N$-body Hamiltonian $H$ 
is well behaved, i.e., the ground state
is well defined thanks to the three-body regulator. 
As we show in
Sec.~\ref{sectrimer}, the three-body regulator produces three-body states that
share many characteristics with the pure three-body Efimov state. 
Pure three-body Efimov states are obtained if the two-body interactions are of
zero range and the hyperradial boundary condition at $R_{123}=0$ is
specified~\cite{BraatenHammerReview}.
Since the hyperradial boundary condition or logarithmic derivative can be
imposed via a delta-function in the hyperradius, we refer to this model as
2bZR+3bZR.

Our work considers the
$N$-body ground state using a novel Monte Carlo approach~\cite{zr15} 
that allows for the
treatment of two-body zero-range interactions.
The Monte Carlo approach can unfortunately not treat three-body zero-range
interactions, i.e., it is not capable of treating the Hamiltonian 2bZR+3bZR.
 A key
objective of the present work is then to investigate how the properties of 
$N$-body droplets in the ground state, supported by the model Hamiltonian 2bZR+3bRp,
change with the number of particles and with the power $p$ of the three-body
regulator.
An important question is to which degree the $N$-body properties are determined
by the three-body parameter.

 For comparison, we
also consider Hamiltonian with finite-range two-body Gaussian or van der Waals
interactions and no three-body interaction. The ground state manifolds of
these models, referred to as 2bG, 2bLJ, 2b10-6, and 2b8-6 (see Table~\ref{tab1}),
lack---as
we show---a number of key Efimov characteristics. Two-body Gaussian
interactions have been
employed extensively in the literature~\cite{javier11,kievsky14,kievsky142,
naidon14,statistics14}, sometimes also in combination with a repulsive
three-body regulator~\cite{javier10,gattobigio11}.

Although the structural properties of the ground state trimers for the
Hamiltonian with two-body van der Waals interactions differ notably from those
for the pure Efimov trimer~\cite{dorner14,dorner15}, these systems exhibit universal features~\cite{
PhysRevLett.107.120401,Pollack18122009,PhysRevLett.103.163202,
PhysRevLett.101.203202,PhysRevLett.102.165302,PhysRevLett.108.145305,
2009NatPh5586Z,greene12,naidon14}.
Specifically, the trimer ground state binding momentum $\kappa_3^{(1)}$ at unitarity is, to a
good approximation, determined by the van der Waals 
length $L_{\text{vdW}}$~\cite{greene12,naidon14} 
and independent of the short-range details.
For the two-body Lenard-Jones potential, one finds
$\kappa_3^{(1)}\approx0.230/L_{\text{vdW}}$~\cite{blume15}, where $L_{\text{vdW}}=(\sqrt{m c_6}/\hbar)^{1/2}/2$. 
This relationship is nowadays being attributed to van der Waals universality. 
Moreover, the binding momentum spacing of $23.4$ between the ground state and the 
first excited state is quite close to the spacing of $22.6944$ exhibited by 
consective pure Efimov trimers~\cite{blume15}. 
It is thus interesting to investigate if 
van der Waals universality exists for $N>3$, i.e., to answer the question 
whether or not the $N$-body ground state energy depends on the short-range 
details of the two-body van der Waals potential.

The remainder of this paper is organized as follows.
Section~\ref{sectrimer} compares the properties of the three-boson system with
infinitely large $s$-wave scattering length interacting through 2bZR+3bZR,
2bZR+3bHC, and 2bZR+3bRp and illustrates the benefits and limitations of these
models. 
Section~\ref{secclusteroverview} reviews several literature results
for $N$-body droplets.
Section~\ref{sec_nbodypimc} extends the calculations for
the 2bZR+3bRp interaction model to clusters with $N\le15$.
In addition to
the energy, various structural properties are discussed in detail.
Section~\ref{secnbodyother}
compares the results for the model 2bZR+3bRp
 with those for systems with two-body
finite-range interactions (i.e., for the models 2bG, 2bLJ, 2b10-6, and 2b8-6).
Finally, Sec.~\ref{secconclusion} concludes.

\section{Three-body system at unitarity}
\label{sectrimer}
To understand the three-body system, it is instructive to rewrite the
Hamiltonian $H$, Eq.~(\ref{hamiltonian}), for $N=3$ in hyperspherical
coordinates~\cite{lin95}.
To this end, we first separate off the center of mass degrees of freedom and
restrict ourselves to states with vanishing relative orbital angular momentum.
For the 2bZR+3bZR, 2bZR+3bHC,
and 2bZR+3bRp models with infinitely large two-body $s$-wave scattering
length $a_s$, the hyperradial and hyperangular degrees of freedom
separate~\cite{BraatenHammerReview,jonsell02}.
The lowest eigen value of the hyperangular
Schr\"odinger equation is typically denoted by
$s_0$, where 
$s_0\approx 1.006 \imath  $~\cite{efimov70,BraatenHammerReview}.
This eigen value enters into the hyperradial 
Schr\"odinger equation with hyperradial Hamiltonian $H_R$,
\begin{equation}
 H_R=-\frac{\hbar^2}{2 m} \frac{\partial^2}{\partial
 R^2}+
 \frac{\hbar^2(s_0^2-1/4)}{2 m R^2}+V_{\text{3b}}(R)
 \label{hyperradialhamiltonian}
\end{equation}
(for notational simplicity, the three-body hyperradius is denoted by $R$
throughout this section).
If $V_{\text{3b}}(R)$ is equal to zero, the eigen energies of the Hamiltonian $H_R$ 
are not well defined.
To make the problem well-defined without explicitly introducing a length scale,
a boundary condition at $R=0$, which serves as a regulator and
defines a scale, can be specified. This is the model 2bZR+3bZR.
The energy spectrum of the 2bZR+3bZR model Hamiltonian displays a perfect 
geometric series~\cite{BraatenHammerReview}.
For an eigen state with binding momentum $\kappa_3^{(n)}$ [the corresponding
  energy is $(\hbar\kappa_3^{(n)})^2/m$], there exists a
tighter and a looser bound state with binding momentum
$\kappa_3^{(n-1)}=\exp(\pi/|s_0|)\kappa_3^{(n)}$
and $\kappa_3^{(n+1)}=\exp(-\pi/|s_0|)\kappa_3^{(n)}$, respectively.
Here, $\exp(\pi/|s_0|)$ is approximately equal to $22.6944$. 
The three-body spectrum for the 2bZR+3bZR model is not bounded from below; in our
notation, this means that $n$ can take non-positive values, i.e., $n=\dots, -2, -1,
0, 1, 2, \dots$
There exists an infinity  of three-body bound states and each hyperradial wavefunction
$\psi_n(R)$ has infinitely many nodes.
The hyperradial wavefunctions of these states collapse if scaled by 
the binding momentum $\kappa_3^{(n)}$, i.e., 
$(\kappa_3^{(n)})^{1/2} \psi_n(\kappa_3^{(n)} R)$ is the same for all states.

\begin{figure}[!tbp]
\centering
\includegraphics[angle=0,width=0.4\textwidth]{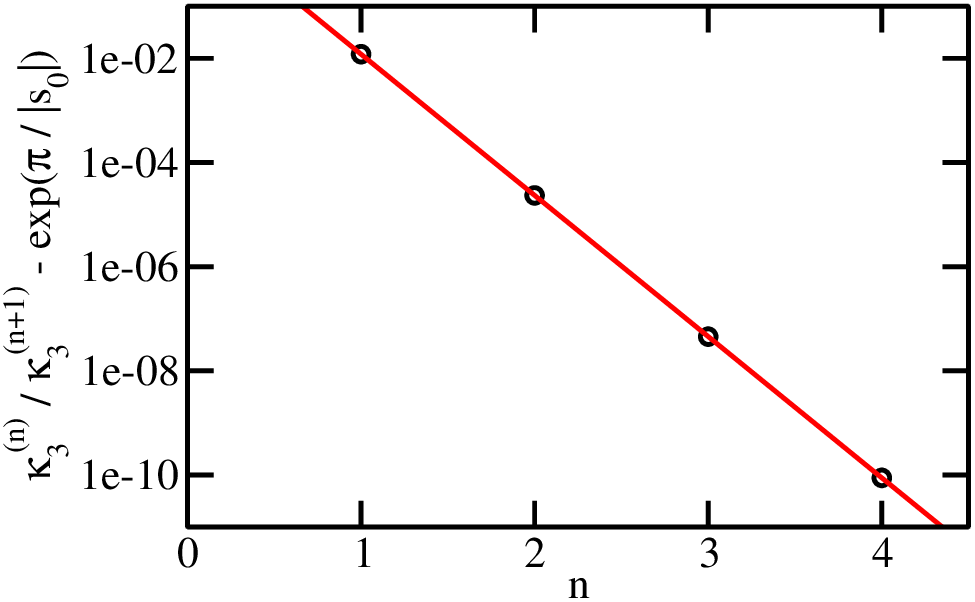}
\caption{(Color online)
  Breaking of the scale invariance for the three-boson 
  system at unitarity with
  three-body hardcore regulator.
  The circles show the difference between the binding momentum ratio
  $\kappa_3^{(n)}/\kappa_3^{(n+1)}$ of the $n$th and $(n+1)$th states for the
  model 2bZR+3bHC and the ratio $\exp(\pi/|s_0|)=22.6944$ for the model
  2bZR+3bZR as a function of $n$.
  The solid line shows a fit to the data points.
  The breaking of the scale invariance becomes weaker with increasing $n$.
 }\label{Fig2}
\end{figure}

We now consider finite-range three-body regulators.
As a first toy model, we consider a hardcore repulsive three-body potential,
i.e., we consider the model 2bZR+3bHC (see Table~\ref{tab1}).
In this case, the hyperangular and hyperradial parts separate as before and
the Hamiltonian $H_R$ supports a well defined ground state with
energy $E_3^{(1)}$  or binding momentum $\kappa_3^{(1)}$ (in our notation,
$n=1,2,\dots$). 
For the $n$th state with binding momentum $\kappa_3^{(n)}$, the hyperradial
wavefunction has $n-1$ nodes.
The circles in Fig.~\ref{Fig2}
show the difference between the 
binding momentum ratios for the model 2bZR+3bHC and the model 2bZR+3bZR.
The binding momentum ratio for the ground and first excited states of the model
2bZR+3bHC is
approximately $22.7064$. The deviation from the model 2bZR+3bZR is $0.0120$
or $0.053\%$. As we go to excited states, the deviations decrease
exponentially. A log-linear fit of the deviations 
yields
$\kappa_3^{(n)}/\kappa_3^{(n+1)}-\exp(\pi/|s_0|)\approx \exp(1.823-6.244n)$ 
(see the solid line in Fig.~\ref{Fig2}).
The overlap between the wavefunction of the ground state of the model
2bZR+3bHC and the wavefunction of the model 2bZR+3bZR with the same
binding momentum is 0.99947, i.e., the inner region where the wavefunction
for the model 2bZR+3bHC deviates from the true Efimov wavefunction
is insignificant.
The three-body hardcore potential breaks the scale-invariance and introduces
$n$-dependent energy spacings.

\begin{figure}[!tbp]
\centering
\includegraphics[angle=0,width=0.4\textwidth]{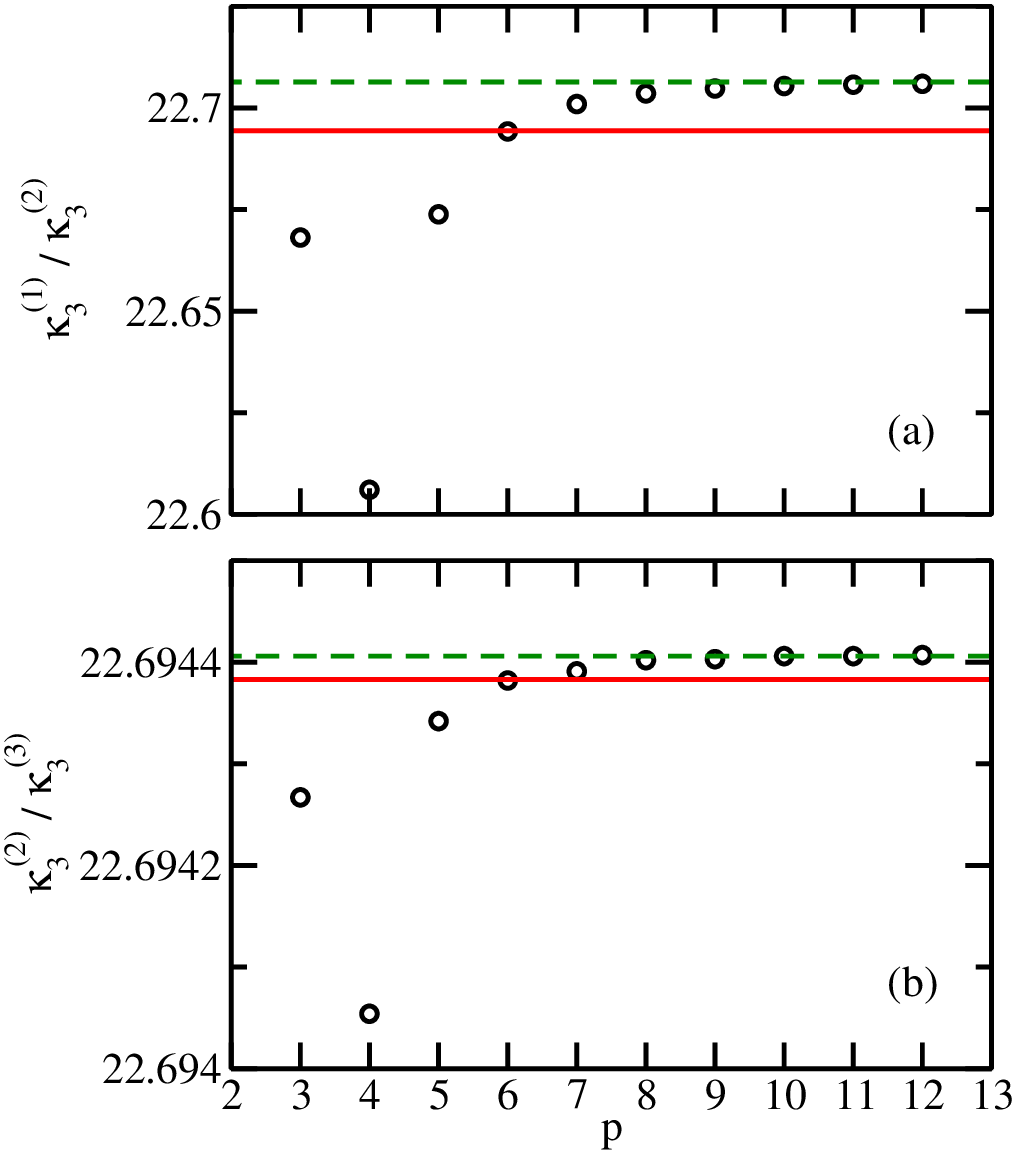}
\caption{(Color online)
  Binding momentum characteristics for the three-boson system with three-body
  power law regulator at unitarity. The circles show the ratio of the binding
  momentum  of two consecutive states for the model 2bZR+3bRp as a function of
  $p$. Panel (a) shows the binding momentum ratio for the ground and the first
  excited states while panel (b) shows the ratio for the first and the second
  excited states. The solid and dashed lines show the binding momentum ratio
  for the models 2bZR+3bZR and 2bZR+3bHC, respectively.
 }\label{Fig3}
\end{figure}

The discontinuity of the derivative of the wavefunction at $R=R_0$
makes the three-body hardcore regulator challenging 
to treat numerically, at least by
the path integral Monte Carlo (PIMC) technique employed in
Sec.~\ref{sec_nbodypimc}. Thus, we consider
three-body power law potentials, which approach the hardcore potential for
$p\to \infty$.
The circles in Fig.~\ref{Fig3}
 show the binding momentum ratios for the model 2bZR+3bRp as a function of $p$.
 Figures 3(a) and 3(b) show the binding momentum ratios for the ground
 and first excited states, and the first and 
 second excited states, respectively.
 As expected, the binding momentum ratios approach the value
 for the model 2bZR+3bHC (dashed lines) in the large $p$ limit.
 For comparison, the solid lines show the binding momentum ratio for the 
 model 2bZR+3bZR.
 The deviations between the binding momentum ratios for the 2bZR+3bRp and the
 2bZR+3bHC models are largest for $p=4$.
 Similar to the model 2bZR+3bHC, the binding momentum ratios for the model
 2bZR+3bRp approach the value $\exp(\pi/|s_0|)$ exponentially with
 increasing $n$.
 
\begin{figure}[!tbp]
\centering
\includegraphics[angle=0,width=0.4\textwidth]{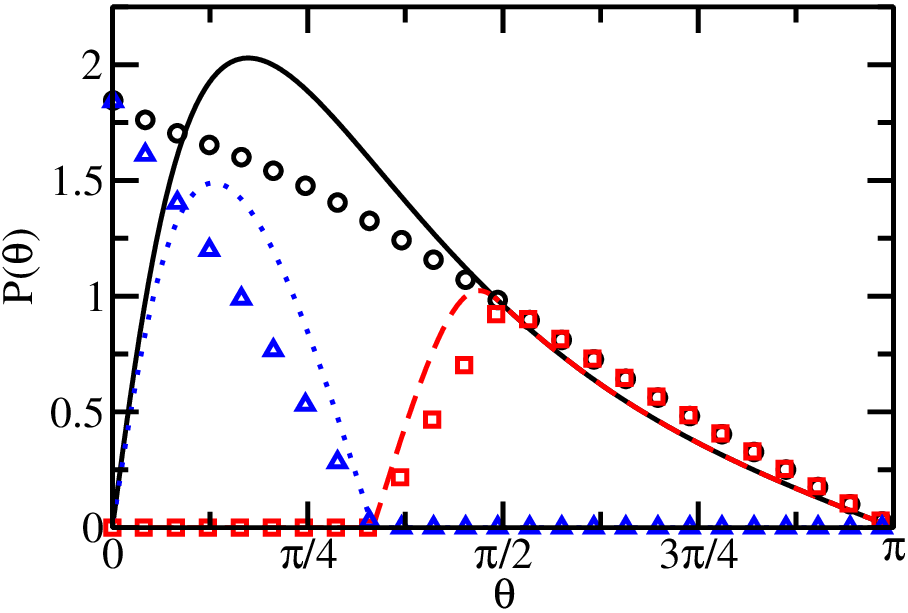}
\caption{(Color online)
Angular distributions for three identical bosons at unitarity. The circles,
triangles and
squares  show the angular distributions $P_{\text{tot}}(\theta)$,
$P_{\text{min}}(\theta)$, and $P_{\text{max}}(\theta)$ for the model 2bZR; these distributions
are identical to those for the models 
2bZR+3bHC and 2bZR+3bRp. The solid, dotted,
and dashed lines show the angular distributions $P_{\text{tot}}(\theta)$,
$P_{\text{min}}(\theta)$, 
and $P_{\text{max}}(\theta)$ for the model 2bG.
 }\label{Fig4}
\end{figure}

 The spacing of the momenta is not the only way to characterize how universal
 the system is, i.e., how close a given system is to the true Efimov scenario
 described by the model 2bZR+3bZR.
 The structural properties provide additional insights.
 Indeed, the structures of weakly-bound three-body systems with positive $a_s$ have
 recently been measured~\cite{dorner14,dorner15}.
 We first look at the distribution of the angles $\theta_{jkl}$ 
 between each pair of position
 vectors,
 $\theta_{jkl}=\arccos(\hat{\mathbf{r}}_{jk}\cdot \hat{\mathbf{r}}_{kl})$.
 The distribution $P_{\text{tot}}(\theta)$ considers all three angles of each
 triangle, while the distribution $P_{\text{min}}(\theta)$ [$P_{\text{max}}(\theta)$]
 considers only the smallest [largest] of the three angles of each
 triangle.
 The normalizations are chosen such that $\int_0^{\pi}P_{\text{tot}}(\theta)d\theta=3$
 and  
 $\int_0^{\pi}P_{\text{min}}(\theta)d\theta=\int_0^{\pi} P_{\text{max}}(\theta)d\theta=1$.
 For infinitely large $a_s$ (as considered throughout this section),
 these angular distributions only depend on the hyperangles and not on the
 hyperradius. Thus,
 they are the same for the models 2bZR+3bZR, 2bZR+3bHC, and 2bZR+3bRp.
 The circles, triangles, and squares in Fig.~\ref{Fig4} show $P_{\text{tot}}(\theta)$,
 $P_{\text{min}}(\theta)$, and $P_{\text{max}}(\theta)$, respectively, for these models.
 $P_{\text{tot}}(\theta)$ is approximately linear and 
 approaches a finite value for $\theta\to0$.
 We are interested in the angular distributions for two reasons.
 (i) For the models 2bG, 2bLJ, 2b10-6, and 2b8-6, the hyperangular and hyperradial 
 degrees of freedom do not separate and the difference between 
 their angular distributions and those for the two-body 
 zero-range models provides valuable insights
 (see Ref.~\cite{blume15}).
 (ii)
 For the $N$-body clusters, the angular distributions, which depend on both the
 hyperangles and the $N$-particle 
 hyperradius, can serve to monitor the three-body
 correlations.

 Solid, dotted, and dashed lines in Fig.~\ref{Fig4}
show the angular distributions $P_{\text{tot}}(\theta)$, $P_{\text{min}}(\theta)$, and
$P_{\text{max}}(\theta)$, respectively, of the three-body 
ground state for the model 2bG.
 Compared to that for 
 the two-body zero-range models, the angular distribution near
 $\theta=0$ for the finite-range model displays distinctly
 different behavior. 
 For the Gaussian model, the probability of finding an angle of zero 
 is  zero and 
 the angular distribution peaks at around $0.17\pi$ or $31^\circ$.
 For the zero-range model, the angular distribution peaks at 0
 and $P_{\text{tot}}(0)$ is finite.
 This is because the zero-range
 boundary condition makes the probability to find two particles 
 at the same position finite.
 A vanishing interparticle distance corresponds  to a triangle 
 in which one of the three angles $\theta_{jkl}$ is zero.
 Since the angular distributions for the models 2bZR+3bZR and 2bG show distinctly
 different
 features, one might expect that the binding momentum ratios
 $\kappa^{(1)}/\kappa^{(2)}$ for these two models also differ.
 The value of $\kappa_3^{(1)}/\kappa_3^{(2)}$
 for the model 2bG is $22.983$, which differs by
 only $1.27\%$ from that
 for the model 2bZR+3bZR.
 This indicates that it is insufficient to only evaluate the binding momentum
 ratios to judge how universal the system is.
 We note that the distribution $P(\theta)$ for the ground state of the $N=3$
 system with two-body Lenard-Jones interactions is quite similar to that for
 the ground state of the $N=3$ system with two-body Gaussian 
interactions~\cite{blume15}.

\begin{figure}[!tbp]
\centering
\includegraphics[angle=0,width=0.4\textwidth]{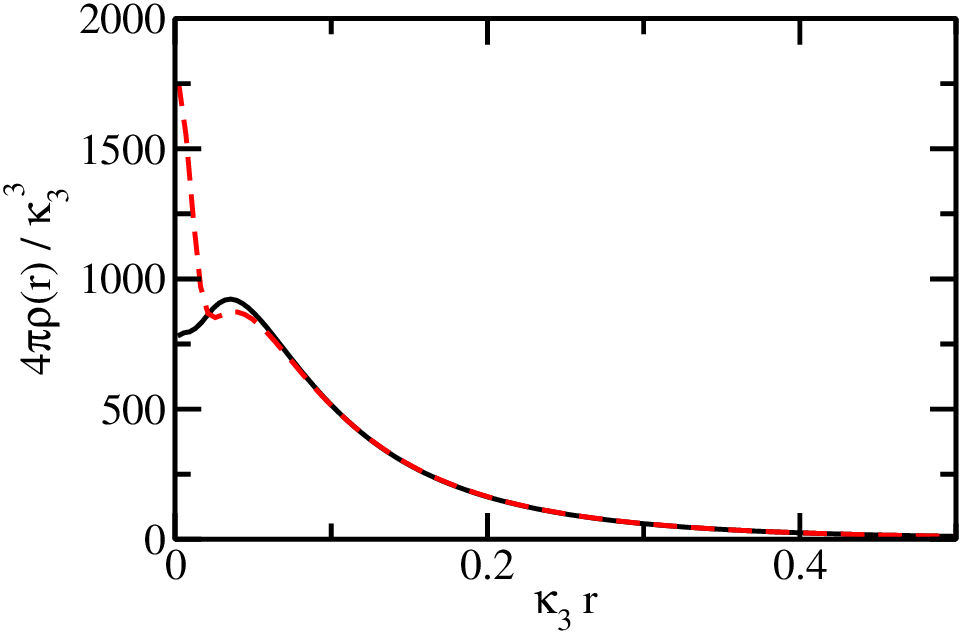}
\caption{(Color online)
  Radial density $\rho(r)$ for three identical bosons at unitarity ($r$ is
  measured relative to the center of mass of the three-body system). The
  dashed and solid lines show $\rho(r)$ for the models 2bZR+3bZR and 
  2bZR+3bRp with $p=6$, respectively.
 }\label{Fig5}
\end{figure}

 We now consider the radial density $\rho(r)$ ($r$ is measured relative to the
 center of mass of the three-body system) for the models 2bZR+3bZR and 2bZR+3bRp
 with  $p=6$.
 The radial density $\rho(r)$ is normalized such that
 $4\pi\int_0^{\infty} \rho(r) r^2dr=N$ and
 depends on the hyperradius and the hyperangles.
 The dashed and solid lines in Fig.~\ref{Fig5}
show the radial density $\rho(r)$ for the models 2bZR+3bZR and
2bZR+3bRp with $p=6$, respectively. 
For the latter, the ground state density is shown.
The radial densities are scaled by their respective binding momentum $\kappa_3$.
The solid and dashed lines agree well in the large $r$ region and 
differ notably in the small $r$ region.
 The deviation in the small $r$ region comes from the fact that the 
 hyperradial density for the model 2bZR+3bZR 
 decays much slower for small $R$ than that for the model 2bZR+3bRp.
 Note that even though the radial densities for the two
 models differ by about 
 a factor of two in the small $r$ region, the difference
 between the integrated contributions is small 
 because the volume element contains an $r^2$ factor.

\section{$N$-body clusters at unitarity: Overview of literature results}
\label{secclusteroverview}
\begin{figure}[!tbp]
\centering
\includegraphics[angle=0,width=0.4\textwidth]{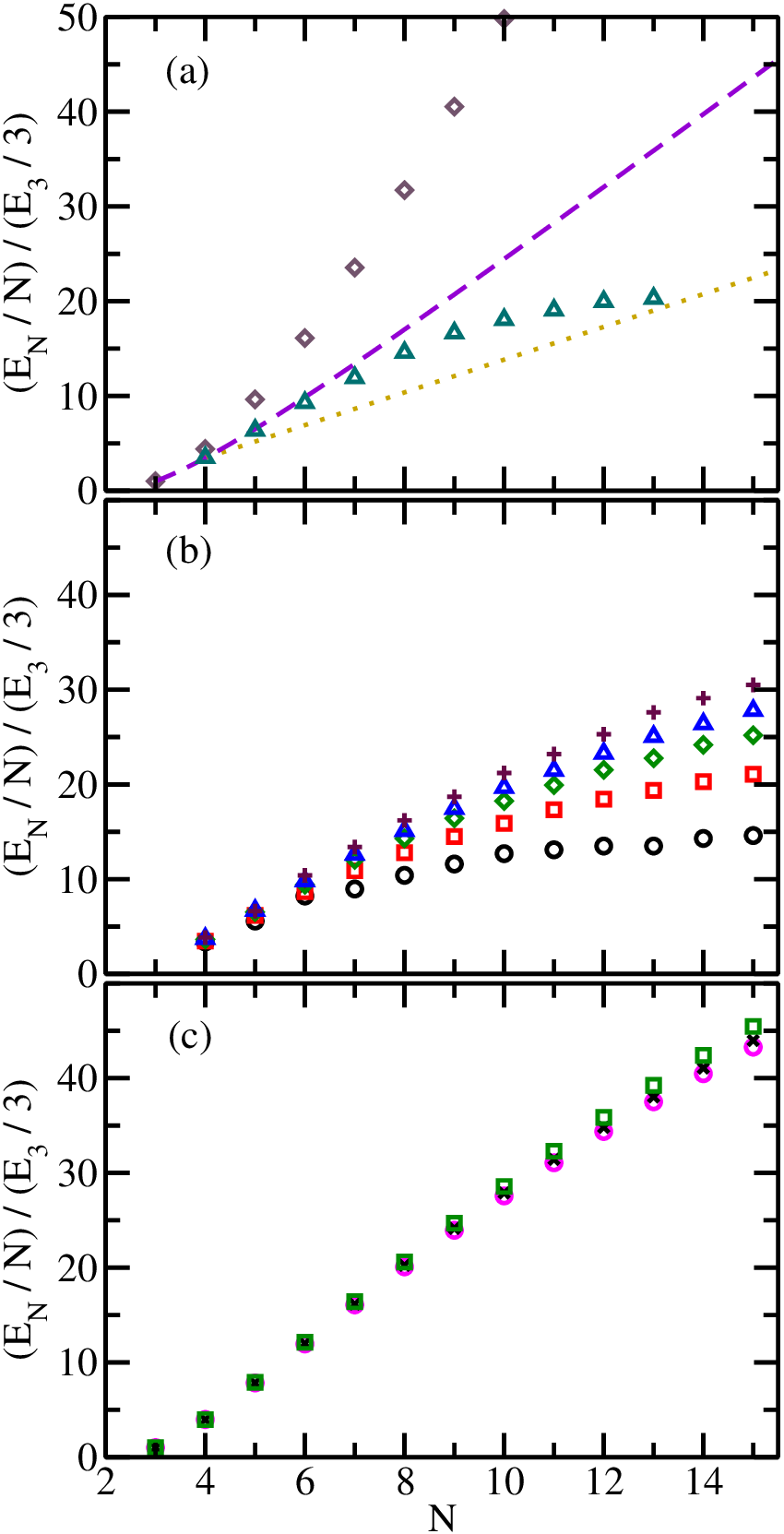}
\caption{(Color online)
  Energy per particle of $N$-boson clusters at unitarity.
  (a) Summary of literature results.
  The dashed and dotted lines 
  show the analytical prediction by Gattobigio and Kievsky~\cite{kievsky14}
  and Nicholson~\cite{amy12}, respectively.
  The triangles show 
  the diffusion Monte Carlo (DMC) energies for a Hamiltonian with two-body square well interaction
  and repulsive three-body hardcore regulator~\cite{javier10}.
  The diamonds show
  the energy for the model 2bG~\cite{statistics14}.
  (b) Summary of our PIMC calculations.
  The circles and pluses are for the model 2bZR+3bRp with $p=4$ and $8$,
  respectively; the error bars (not shown) are of the order of the symbol sizes.
  The squares, diamonds, and triangles are for the 
  model 2bZR+3bRp with $p=5,6$, and $7$,
  respectively; the error bars (not shown) are smaller than the symbol sizes.
  (c) Summary of our calculations for two-body van der Waals models.
  The circles, crosses, and squares show our DMC results for the models 2bLJ,
  2b10-6, and 2b8-6, respectively.
 }\label{Fig6}
\end{figure}

This section discusses various 
literature results for the energy of weakly-bound $N$-body droplets 
($N>3$) at unitarity.
The diamonds in Fig.~\ref{Fig6}(a)
show the $N$-boson energy per particle $E_N/N$ for the 
model 2bG as a function of $N$~\cite{kievsky14,kievsky142,statistics14}.
The energy per particle increases approximately linearly with $N$ for $N>6$
(for smaller $N$, some deviations from the linear behavior exist).
Based on the fact that the energy per particle, and correspondingly
the binding momentum, scale linearly with $N$ for
the model 2bG, 
Gattobigio
\textit{et al.}~\cite{kievsky14} proposed 
an analytical form for the $N$-boson system with two-body zero-range
interactions and fixed three-body parameter,
\begin{eqnarray}
\label{eq_kappa}
\frac{\kappa_N}{\kappa_3}=1+ \left( \frac{\kappa_4}{\kappa_3}-1 \right)(N-3)
\end{eqnarray} 
[see the dashed line in
  Fig.~\ref{Fig6}(a)].
The ratio $\kappa_4/\kappa_3$ is not taken from the ground state calculations
for the Gaussian two-body interaction model, for which
$\kappa_4/\kappa_3=\sqrt{5.86}$, but from Deltuva's calculations for highly
excited four-body resonance states. Deltuva finds
the universal ratio
$\kappa_4/\kappa_3=\sqrt{4.61}$~\cite{deltuva10}. Gattobigio \textit{et al.}'s
expression, converted to the energy, exhibits a leading order $N^2$ and
sub-leading order $N$ dependence.

It should be noted that the ground state energy of 
the Hamiltonian with
pairwise Gaussian interactions scales differently with $N$ for
$N\gtrsim 10$ than
that of Hamiltonian with pairwise interactions with short-range repulsion.
For interactions with repulsive core,
it is well-established that the energy per particle
increases weaker than linear for $N\gtrsim 10$ (see, e.g., the literature
on helium and tritium 
droplets~\cite{whaley94,hanna02}).
Gattobigio {\em{et al.}}~\cite{kievsky142} 
noted that Eq.~(\ref{eq_kappa}) applies not only 
to systems with zero-range interactions but also
to systems with finite-range interactions in the regime where 
$E/N$ is approximately proportional
to $N$ 
(e.g., to helium droplets with $N \lesssim 10$). 
In this case, the ratio $\kappa_4/\kappa_3$ for the
finite-range potential is taken as input and the binding momentum for 
$N>4$ is predicted.
We return to this discussion in Sec.~\ref{secnbodyother}.

Independent evidence for the leading-order $N$ dependence of the 
energy per particle for the Hamiltonian with two-body zero-range interactions
comes from lattice calculations
for even $N$~\cite{amy12}.
Assuming that 
the distribution of the two-body correlator is exactly
log normal, Nicholson deduced an analytical expression for the energy per
particle,
$E_N/N=(N/2-1)E_4/4$ [see the dotted line in Fig.~\ref{Fig6}(a)]~\cite{amy12}.
To plot this expression, we used Deltuva's value of $E_4/E_3=4.61$.
It should be noted that the coefficients predicted by Gattobigio \textit{et
al.} and Nicholson for the leading order $N$ dependence differ by about a
factor of $2$.

A somewhat different $N$-dependence of the energy per particle 
was observed in the
numerical calculations by von Stecher [see triangles in
Fig.~\ref{Fig6}(a)]~\cite{javier10}.
In fact,
the idea to use a three-body regulator, as in our model 2bZR+3bRp, to make the
ground state trimer large and Efimov-like was introduced in
Ref.~\cite{javier10}.
Von Stecher employed a model Hamiltonian with 
two-body square well potential with infinitely large 
two-body $s$-wave scattering
length and three-body hardcore potential.
For $N\lesssim 10$, the energy per particle increases approximately
linearly with increasing $N$.
For larger $N$, the triangles in Fig.~\ref{Fig6}(a) flatten.
Reference~\cite{krauth13} interpreted this as a turnover to a $N^0$
dependence of the energy per particle.
Such a behavior suggests a saturation of the density for large $N$.
This saturation would be a consequence of 
the balance of the two-body attractive
and three-body repulsive interactions.

The discussion above shows that the dependence of the energies tied to Efimov
trimers is not well understood. Specifically, neither the functional form of
the energy per particle nor
the coefficients are agreed upon. In the following
sections, we attempt to understand
where the discrepancies of the literature results come from.

\section{$N$-body results at unitarity for the model $\text{2bZR+3bRp}$}
\label{sec_nbodypimc}
To calculate the $N$-boson energy for the Hamiltonian with interaction model
2bZR+3bRp,
we apply the PIMC technique~\cite{ceperleyrev,zr15}.
The PIMC technique is an, in principle, 
exact finite-temperature method; the errors, which originate
from the discretization of the imaginary time and the stochastic evaluation of
integrals, can be reduced systematically.
To obtain the ground state energy of the $N$-boson Hamiltonian,
the PIMC approach has to be extended to the zero-temperature limit.
Typically, this is achieved by the path integral ground state 
approach~\cite{ceperleyrev,pigs99}. Here,
we pursue an alternative strategy. Namely, we work in the finite temperature
regime where the thermal contribution to the energy is known and where the
structural properties of interest are not affected by the temperature. This
approach was introduced and benchmarked in Ref.~\cite{statistics14}. The basic
idea is to place the droplet in a weak external harmonic confinement, whose
angular frequency $\omega$ is chosen such that the center of mass energy
spectrum becomes discretized and the relative motion is unaffected by the
trap. This requirement corresponds to $|E_N|\gg \hbar\omega$.
Since the density of states of the harmonically trapped center of
mass pseudoparticle is known analytically, the ground state energy $E_N$ of
the $N$-boson droplet in free space can be extracted from the finite-temperature
energy~\cite{statistics14,zr15}.

The circles, squares, diamonds, triangles, and pluses in Fig.~\ref{Fig6}(b)
show the
energy per particle for the model 2bZR+3bRp with $p=4, 5, 6, 7,$ and $8$,
respectively, as a function of $N$
 (see also Table~\ref{tabpimc}
 and the Supplemental Material~\footnote{The Supplemental Material at [To Be Inset
 by the editor]
 contains tables for the energies 
 for the models 2bZR+3bRp with $p=4-8$, He-He(scale), and He-He(arctan).}).
\begin{table}[!tbp]
  \centering
  \caption{
    PIMC energies for the model 2bZR+3bRp for $N=4-15$. Columns 2-4 show the
    scaled energy 
    $E_N/N / (E_3/3)$ for $p=5,6,$ and $7$, respectively. The error bars (not
    explicitly reported) are around 3\%.
  }
\begin{tabular} 
{c c c c}
    \hline
    \hline
    $N$& 2bZR+3bR5 & 2bZR+3bR6 & 2bZR+3bR7 \\
    \hline
4 &  3.46 &   3.64 & 3.73  \\
5 &  6.19 &   6.53 & 6.70  \\
6 &  8.69 &   9.42 & 9.81  \\
7 &  10.9 &   12.0 & 12.6  \\
8 &  12.8 &   14.3 & 15.1  \\
9 &  14.5 &   16.4 & 17.5  \\
10&  15.9 &   18.3 & 19.7  \\
11&  17.3 &   20.0 & 21.5  \\
12&  18.4 &   21.5 & 23.3  \\
13&  19.4 &   22.8 & 25.0  \\
14&  20.3 &   24.2 & 26.4  \\
15&  21.1 &   25.2 & 27.8  \\
    \hline
    \hline
  \end{tabular}
  \label{tabpimc}
\end{table}
For each $p$, the energy per particle is scaled by the respective trimer energy per particle.
For a fixed $p$, the energy per particle increases monotonically and smoothly
as a function of $N$, i.e, even-odd effects, which have been observed 
in trapped and homogeneous two-component Fermi gases~\cite{carlson03,
chang07,blume07}, are---if existent---smaller than our statistical
error bars.
For fixed $N$, the scaled energy per particle increases with increasing $p$
($p\ge 4$); this increase becomes smaller with increasing $p$.
Similarly to von Stecher's energy per particle~\cite{javier10}
[triangles in Fig.~\ref{Fig6}(a)],
the scaled energy per particle increases roughly linearly for
smallish $N$
and then flattens out for larger $N$.
This effect is most pronounced for $p=4$ and $5$, where the flattening sets in
around $N=8-10$, and least pronounced for $p=8$.
The reason for the flattening is that the clusters develop, for sufficiently large $N$,
more than one pair distance scale (see below for more details).

\begin{figure}[!tbp]
\centering
\includegraphics[angle=0,width=0.4\textwidth]{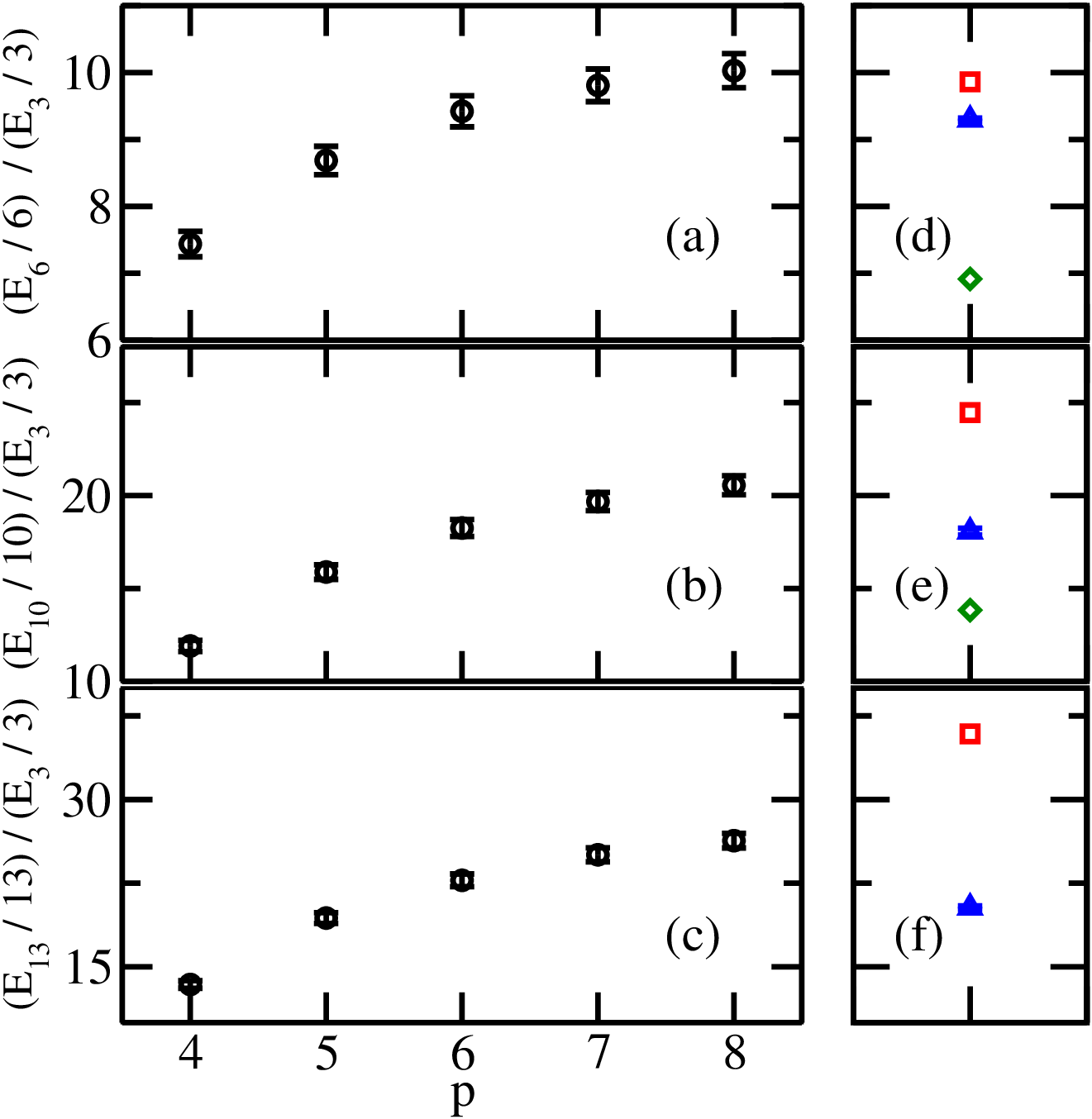}
\caption{(Color online)
  Comparison of our PIMC energies (left) and literature results (right) for
  selected $N$.
  Panels (a), (b), and (c) show our PIMC energy per particle for
  $N$-boson clusters interacting through 
  the model 2bZR+3bRp as a function of $p$
  for $N=6, 10,$ and $13$, respectively.
  For comparison,
  panels (d), (e), and (f) show the energy per particle from the literature for the same
  $N$.
  The triangles, diamonds, and squares are from von Stecher~\cite{javier10}, 
  Nicholson~\cite{amy12}, and Gattobigio \textit{et al.}~\cite{kievsky14}.
  Since the work by Nicholson is restricted to even $N$, comparison for
  $N=13$ cannot be made.
 }\label{Fig7}
\end{figure}

The circles in Fig.~\ref{Fig7} 
replot the PIMC 
energy per particle for selected $N$. As the power $p$ increases,
the scaled
energy approaches a constant. Based on our discussion in Sec.~\ref{sectrimer},
the $p\to\infty$ energy should coincide with the energy for
the model 2bZR+3bHC. 
It is thus instructive to compare our scaled energies, extrapolated by eye
to the
$p\to \infty$ limit, with those obtained by von Stecher~\cite{javier10}, 
who employed a
two-body square well potential and a
three-body hardcore regulator [see
  triangles in Figs.~\ref{Fig7}(d)-\ref{Fig7}(f)]. 
  We find that our $p\to \infty$
  energy per particle lies above von Stecher's energy per particle by
  something like 
  $10-20\%$, $20-30\%$, and $30-50\%$ for $N=6,
  10,$ and $13$, respectively.
  Since the three-body sectors are treated on consistent footing (3bRp$\to$3bHC
  as $p\to\infty$), we speculate that the difference arises from the different
  two-body interactions.
  However, we did not perform calculations to confirm this and can thus
  not rule out other reasons. As can be seen from Fig.~\ref{Fig7}, Nicholson's
  energy prediction lies notably below our large $p$ energies while Gattobigio
  \textit{et al.}'s prediction lies above our large $p$ energies for
  $N\gtrsim 8$.

If the $N$-body energies were determined solely by a three-body
parameter $\kappa_3$, the model 2bZR+3bRp
for different $p$ would yield the same scaled energies, i.e., the symbols in
Fig.~\ref{Fig6}(b) would collapse to a single curve.
The fact that they do not collapse indicates that the three-body parameter is
not sufficient to predict the energy of the $N$-boson clusters, at least not
for the models considered.
To gain more 
insight into this, it is instructive to analyze the length scales of
the model 2bZR+3bRp. Four length scales can be identified (see rows 3--6 of
Table~\ref{tab2}).
\begin{table*}[!tbp]
  \centering
  \caption{
Summary of the definitions of length scales.
The van der Waals length $L_{\text{vdW}}$ is defined in Ref.~\cite{eita10}. $L_p$ for
$p=6$ agrees with $L_{\text{vdW}}$ if $m$ is replaced by the reduced two-body mass
$m/2$.
}
\begin{tabular} 
{c c c}
    \hline
    \hline
     length scale & definition & description        \\
    \hline
$L_g$ &$r_0$ &characteristic length scale of
the two-body Gaussian potential  \\
$L_{\text{vdW}}$ &$(\sqrt{m c_6}/\hbar)^{1/2}/2$ &characteristic
length scale of the two-body van der Waals potential  \\
$L_p$ &$[1/(p-2)\sqrt{2mC_p}/\hbar]^{2/(p-2)}$ &characteristic
length scale of the three-body repulsive potential  \\
$\bar{L}_{3}$ &$1/\kappa_3=\hbar/\sqrt{m |E_3|}$ & length scale set
by the three-body binding energy\\
$\bar{L}_{N}$ &$1/\kappa_N=\hbar/\sqrt{m |E_N|}$  & length scale  set
by the $N$-body binding energy\\
$\bar{l}_{N}$ &$\hbar/\sqrt{m|E_N|/N}=\sqrt{N}\bar{L}_{N}$ & length scale set
by the $N$-body binding energy per particle\\
$\bar{r}$ & & average interparticle spacing\\
$\bar{R}$ & & average sub-three-body hyperradius \\
    \hline
    \hline
  \end{tabular}
  \label{tab2}
\end{table*}
(i) The characteristic length scale $L_p$ of the three-body repulsive potential.
 (ii) The length scale $\bar{L}_{3}$ defined by the three-body binding energy.
 (iii) The length scale $\bar{L}_{N}$ defined by the energy of the cluster.
 And, (iv) the length scale $\bar{l}_{N}$ associated with the energy per
 particle of the cluster.
 Inspection of the definitions given in Table~\ref{tab2} shows that 
 $\bar{L}_{N}$ and $\bar{l}_{N}$ are not independent.

 For $p=4$--$8$, we
 find $\bar{L}_3/L_p\approx29.3, 28.8, 27.6, 26.6,$ and $25.9$,
 i.e., the trimer is significantly larger than
 the scale of the underlying repulsive three-body potential. This ensures, as
 discussed in Sec.~\ref{sectrimer}, that the trimer ground state described by
 the model 2bZR+3bRp with $p\ge4$ exhibits the key characteristics of an
 Efimov state. It is instructive to alternatively think about the trimer size
 in terms of the average interparticle spacing $\bar{r}$. For trimers
 with $p=4$--$8$, we find $\bar{r}/L_p\approx 18.7, 18.5, 17.7$, $17.1,$
 and $16.6$.

 For $p=6$,
 we find that
 $\bar{L}_{N}/ L_p$ changes from $11.2$ for $N=4$ to $8.37$
 for $N=5$ to $2.46$ for $N=15$. This
 suggests that the $N$-boson droplet ``sees'' increasingly more of the
 three-body regulator as $N$ increases, i.e., that the dependence of $E_N/N$
 on $p$ increases with increasing $N$. 
 The length scale
 $\bar{l}_{N}$, 
 in contrast, suggests a larger separation of scales; for $N=13$, e.g.,
 we have
 $\bar{l}_{N}/L_p =7.69$ for $p=4$ and $\bar{l}_{N}/L_p=6.85$ for $p=8$.
 In fact, if $E_N/N$
 scales as $N$,
 then $\bar{L}_{N}$ and $\bar{l}_{N}$ scale as $1/N$
 and $1/\sqrt{N}$, respectively. If $E_N/N$ scales as $N^0$, then 
 $\bar{L}_N$ and $\bar{l}_N$ scale as $1/\sqrt{N}$ and $N^0$, respectively.
 This implies that---unless the energy scales linearly (or even weaker) with
 $N$ for large $N$---the properties of the $N$-boson droplets are expected
 to be notably affected by the choice of the three-body regulator.

\begin{figure}[!tbp]
\centering
\includegraphics[angle=0,width=0.4\textwidth]{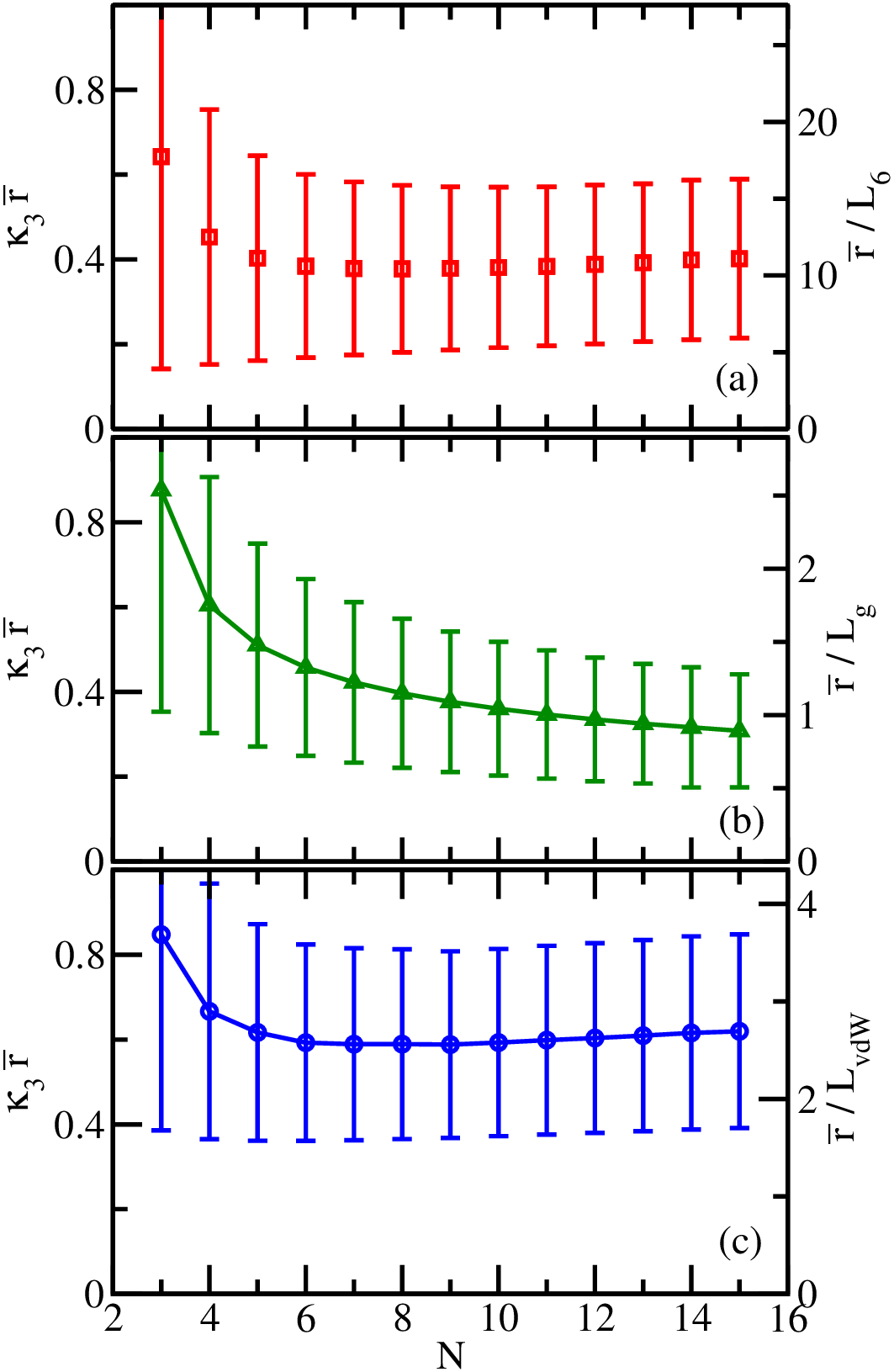}
\caption{(Color online)
  Expectation value $\bar{r}$ of the
  pair distance as a function of $N$ for $N$-boson systems interacting through
  various models.
  (a) The squares are for the model 2bZR+3bRp with $p=6$.
  (b) The triangles are for the model 2bG.
  (c) The circles are for the model 2bLJ.
  The error bars show the variance of the pair distance. 
  The pair distances are plotted using two different units: (i) the
  inverse three-body binding momentum (left axis) and 
  (ii) the characteristic length scale of the model Hamiltonian (right axis).
 }\label{Fig8}
\end{figure}

\begin{figure}[!tbp]
\centering
\includegraphics[angle=0,width=0.4\textwidth]{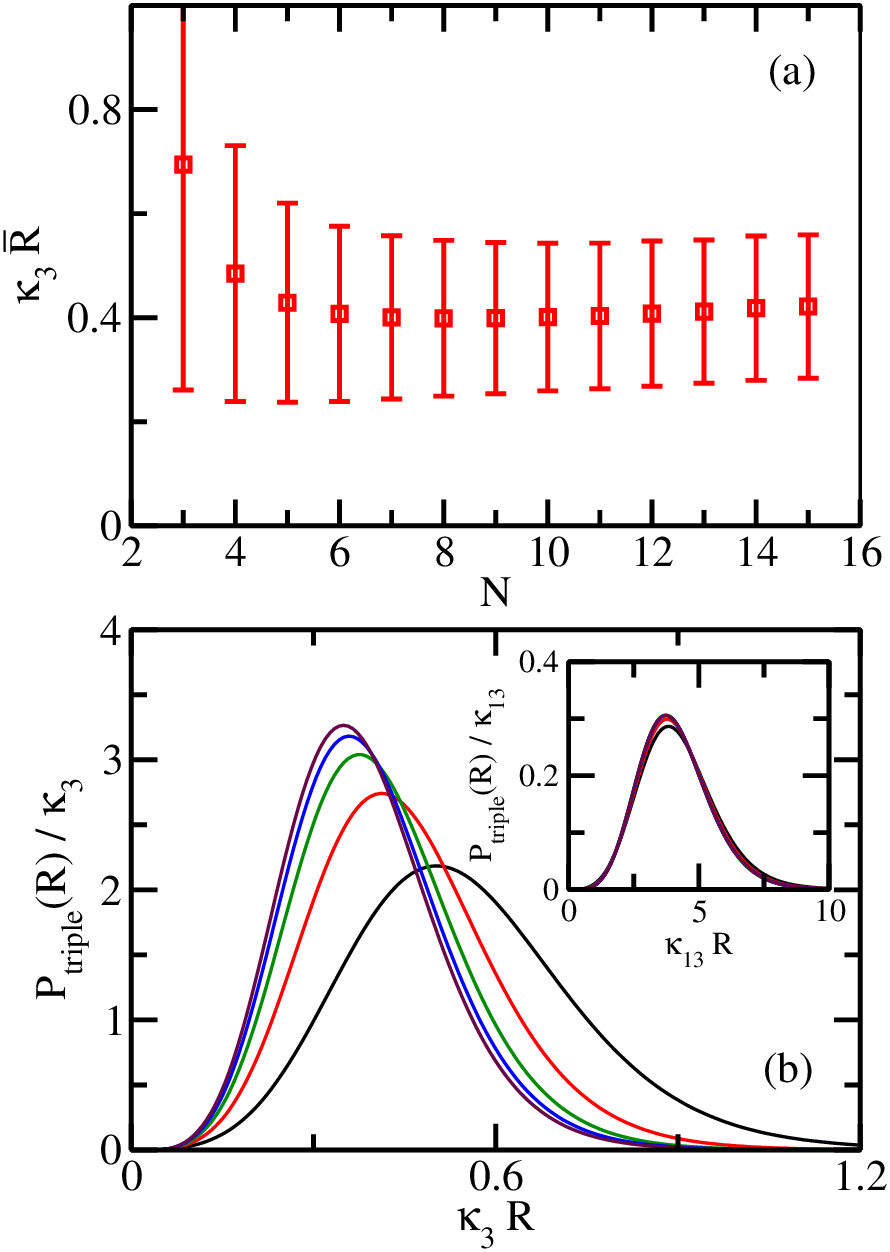}
\caption{(Color online)
  (a) Expectation value $\bar{R}$ of the
  sub-three-body hyperradius (triple size) 
  as a function of $N$ for $N$-boson systems interacting through
  the model 2bZR+3bR6.
  The error bars show the variance of the triple size.
  (b)
  Triple distribution function $P_{\text{triple}}(R)$ for the $N=13$ cluster
scaled using the three-body binding momentum $\kappa_3$.
  The solid lines from top to bottom at $\kappa_3 R=0.6$ are for the model
  2bZR+3bRp with $p=4, 5, 6, 7,$ and $8$.
  The inset replots the triple distribution functions using the 
binding momentum $\kappa_{13}$ of the $N=13$ droplet.
In these units, the triple distribution functions for different $p$ collapse.
}\label{Fig9}
\end{figure}

 Alternatively, one can consider the average interparticle distance
 $\bar{r}$ and the average sub-three-body hyperradius $\bar{R}$.
The squares in Fig.~\ref{Fig8}(a)
show the average interparticle spacing $\bar{r}$, i.e., the 
expectation value of the pair distance, as a function of $N$ 
in units of $1/\kappa_3$ (left axis) and in units of $L_6$ (right axis)
for the model 2bZR+3bRp with $p=6$.
The error bars indicate the variance $\Delta r$
of the pair
distance, $\Delta r = \sqrt{\langle r^2\rangle-\langle r\rangle^2}$, where
$\langle \rangle$ indicates the quantum mechanical expectation
value~\footnote{In practice, the structural properties are obtained by
calculating thermal averages at low temperature, where the excitations of the
relative degrees of freedom are negligible. This implies that thermally averaged
structural properties that are independent of the center of mass degrees of
freedom coincide, to a very good approximation, with the corresponding quantum
mechanical expectation values with respect to the ground state wave function.}.
As the number of particles $N$ increases, both the mean and variance of the
pair distance are nearly constant.
The mean and variance of the pair
distance are about one order of magnitude larger than the internal length scale
$L_p$.
The relatively large variance of the Hamiltonian with model interaction
2bZR+3bRp implies that the clusters are diffuse and liquid-like.
The squares in Fig.~\ref{Fig9}(a)
show the average sub-three-body hyperradius $\bar{R}$, i.e., the 
expectation value of the triple size, as a function of $N$
for the model 2bZR+3bRp with $p=6$. The error bars indicate the variance.
The mean and variance of the
sub-three-body hyperradius behave similar to the mean and variance 
of the pair
distance.

\begin{figure}[!tbp]
\centering
\includegraphics[angle=0,width=0.4\textwidth]{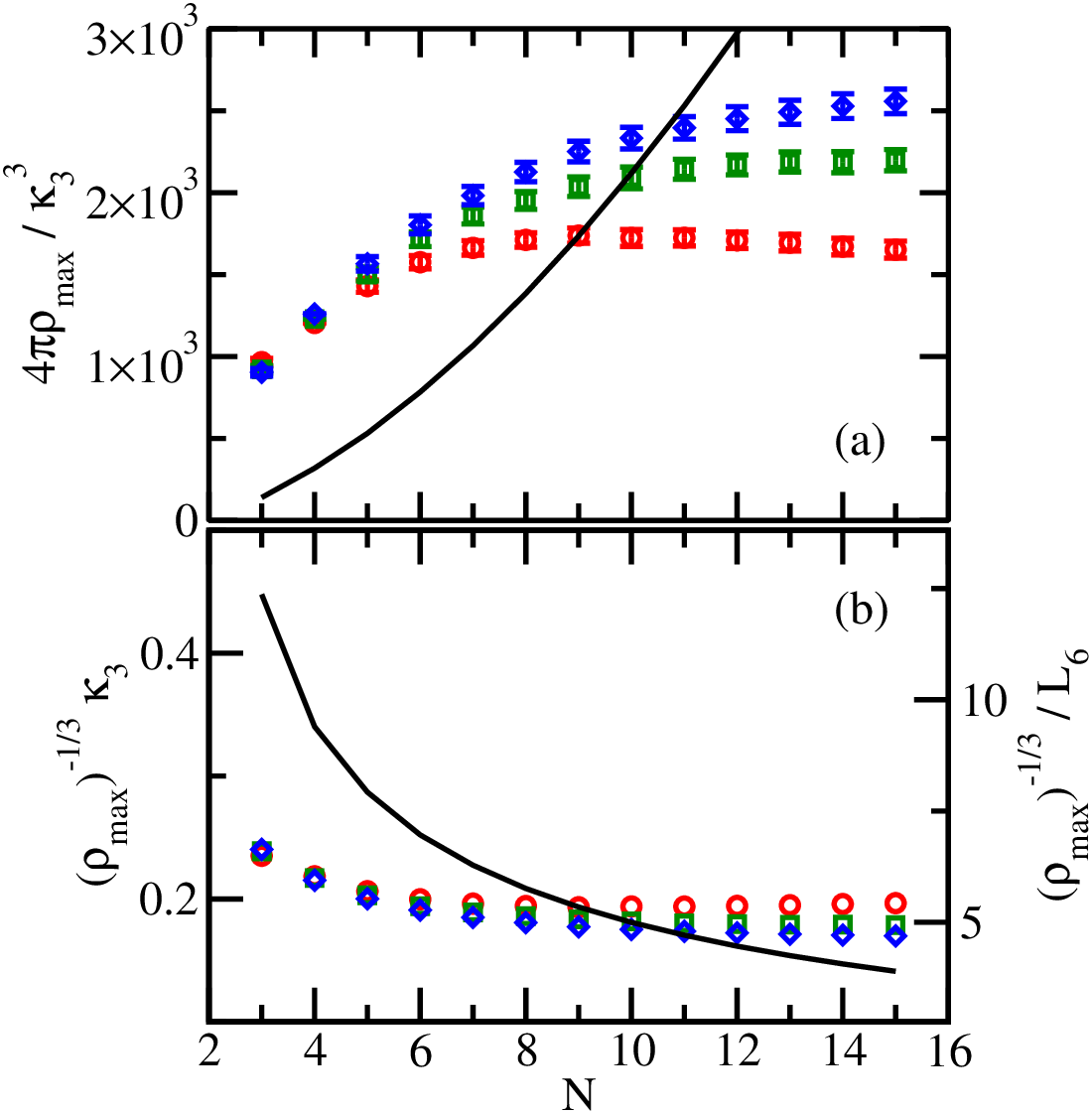}
\caption{(Color online)
  (a) Maximum density $\rho_{\text{max}}$ as a function of $N$ for $N$-boson systems
  interacting through various models.
  The circles, squares, and diamonds are
  for the model 2bZR+3bRp with $p=5$ (lowest data set), $6,$
  and $7$ (highest data set), respectively.
  For comparison, the line is for the model 2bG.
  (b) Same data as in (a) but replotted as the minimum average 
  interparticle distance $(\rho_{\text{max}})^{-1/3}$.
  The right axis shows the data for the model 2bZR+3bR6 in units of $L_6$.
 }\label{Fig10}
\end{figure}

The average pair distance and sub-three-body hyperradius are obtained by
averaging over all
possible pairs and triples regardless of whether or not the particles are
close to each other.
To get more ``local'' information, we calculate the maximum density
and subsequently the 
closest pair distance.
The circles, squares, and diamonds in Fig.~\ref{Fig10}(a)
show the maximum $\rho_{\text{max}}$ of the radial density
for the model 2bZR+3bRp with $p=5,6,$ and $7$, respectively,
as a function of $N$.
We find that unlike for $N=3$ (see Fig.~\ref{Fig5}), the radial density peaks
at $r=0$ for $N\ge4$.
For all $p$, the maximum of the radial density is roughly a constant for the
largest $N$ considered. This constant depends---as the energy per particle---on 
the three-body regulator.
The circles, squares, and diamonds in Fig.~\ref{Fig10}(b)
show the smallest average pair distance 
for the model 2bZR+3bRp with $p=5, 6,$ and $7$, respectively, as a function of $N$.
The smallest average pair distance decreases with increasing $N$
and approximately saturates for the largest $N$ considered. 
The smallest average pair distance is only about five 
times larger than the characteristic length scale $L_p$ of the three-body regulator.

\begin{figure}[!tbp]
\centering
\includegraphics[angle=0,width=0.4\textwidth]{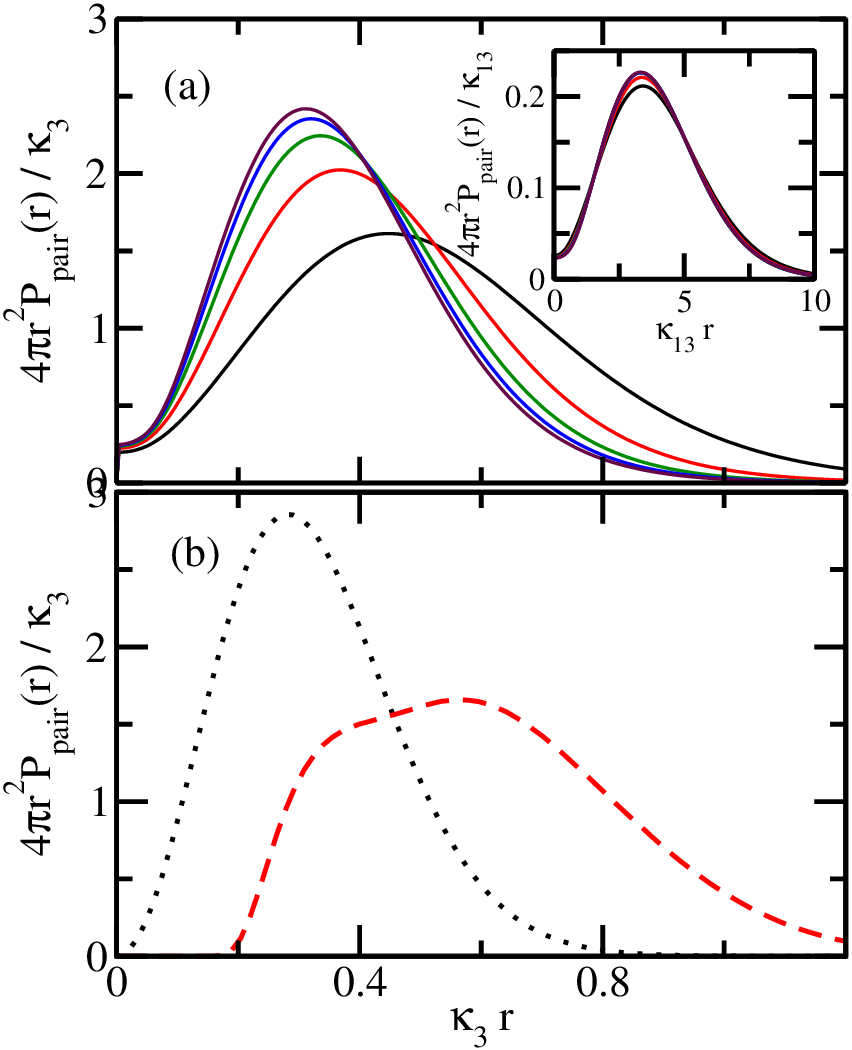}
\caption{(Color online)
  Scaled pair distribution function $4\pi r^2 P_{\text{pair}}(r)$
  for $N=13$ bosons interacting through 
  various models.
  (a) The solid lines from top to bottom at $\kappa_3 r=0.8$ are for
  the model 2bZR+3bRp with $p=4$--$8$,
scaled using the 
three-body binding momentum $\kappa_3$.
  The inset replots the pair distribution functions scaled using the 
binding momentum $\kappa_{13}$ of the $N=13$ droplet.
In these units, the pair distribution functions for different $p$ collapse.
  (b) The dashed and dotted lines
show the scaled distribution functions 
   for the models 2bLJ and 2bG, respectively, using the three-body 
binding momentum $\kappa_3$.
 }\label{Fig11}
\end{figure}

The above length scale discussion can be expanded by considering distribution
functions.
The scaled pair distribution function $4\pi r^2 P_{\text{pair}}(r)$, normalized
according to $4\pi\int_0^{\infty} r^2P_{\text{pair}}(r)dr=1$, tells one the
probability to find two particles at a distance $r$ from each other.
The lines from top to bottom at $\kappa_3 r=0.8$ in Fig.~\ref{Fig11}(a)
show the scaled pair distribution function $4 \pi r^2P_{\text{pair}}(r)$ 
for $N=13$ 
interacting through
2bZR+3bRp with $p=4$--$8$.
The amplitude at $r=0$ is finite and roughly independent of $p$.
This makes sense as it
is a 
signature of the two-body zero-range interactions,
which enforce a finite amplitude at $r=0$.

The triple distribution function $P_{\text{triple}}(R)$, normalized according
to $\int_0^{\infty} P_{\text{triple}}(R)dR=1$, tells one the
probability to find three particles with
sub-three-body hyperradius $R$.
  The solid lines from top to bottom at $\kappa_3 R=0.6$ in
  Fig.~\ref{Fig9}(b)
show the triple distribution function $P_{\text{triple}}(R)$ for $N=13$ interacting
through
2bZR+3bRp with $p=4$--$8$.
The triple distribution functions are broad and structureless,
indicating that the clusters are diffuse and liquid-like and that 
no small three-body sub-systems are formed.

Figures~\ref{Fig9}(b) and  \ref{Fig11}(a) show that the distribution functions
$P_{\text{pair}}(r)$ and $P_{\text{triple}}(R)$ do not collapse if scaled by the three-body binding
momentum $\kappa_3$. 
The distribution functions for
$p=4$ are notably broader than those for $p>4$.
Figures~\ref{Fig9}(a) and~\ref{Fig11}(a) suggest that the distribution functions
converge in the large $p$ limit (i.e., in the three-body hardcore regulator
limit). Similar behavior is observed for other $N$.
As shown in the insets of
Figs.~\ref{Fig9}(b) and  \ref{Fig11}(a), the distribution
functions collapse 
to a very good approximation to a single curve if scaled by the binding
momentum $\kappa_N$ of the $N$-body droplet.
This can be understood as a new type of
universality, which is weaker than the 
``Efimov universality'': The binding momentum
$\kappa_N$ allows one to collapse the distribution functions
for the models 2bZR+3bRp for sufficiently large $p$ but
$\kappa_N$ is not determined by $\kappa_3$ (the latter would 
constitute
``Efimov universality''). 
The dominance of $\kappa_N$ arises because the vast majority
of the wave function amplitude is located in the classically
forbidden region~\cite{jensen04} (for pure zero-range interactions,
the classically allowed region is reduced to a single
point).

\begin{figure}[!tbp]
\centering
\includegraphics[angle=0,width=0.4\textwidth]{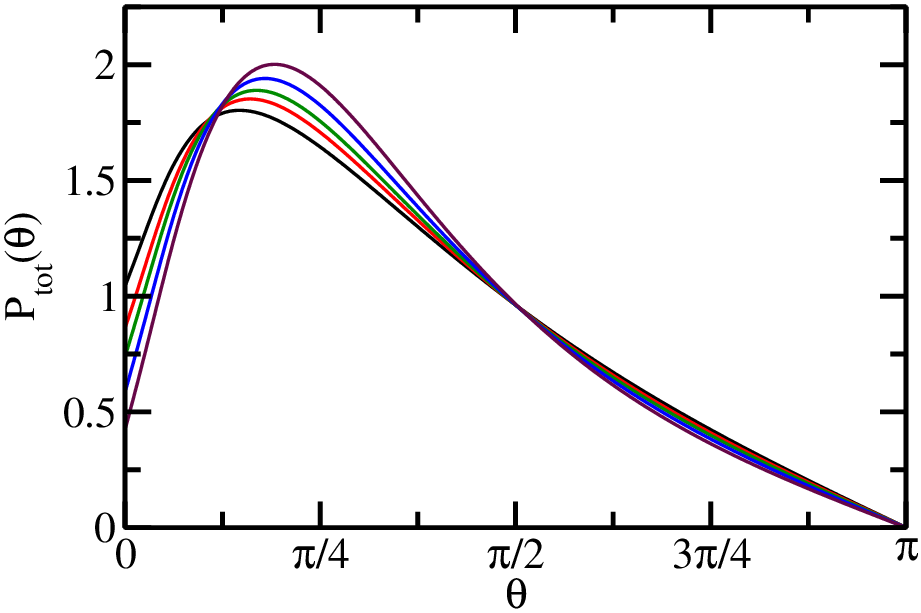}
\caption{(Color online)
  Angular distribution $P_{\text{tot}}(\theta)$ for $N$-boson clusters
  interacting through the 
  model 2bZR+3bRp with $p=6$.
  The lines from top to bottom at $\theta=0$ are for $N=5, 6, 7, 9,$ and
  $13$.
 }\label{Fig12}
\end{figure}

At the three-body level, the angular distribution
functions for the models 2bZR+3bRp
and 2bZR+3bZR coincide since the hyperradial and hyperangular degrees of freedom
separate. This is not the case for
$N>3$, since the three-body 
regulator depends on the $N$-body hyperradius and a subset of the $3N-4$
hyperangles.
For fixed $N$, we find that the dependence of the angular distribution
functions
$P_{\text{tot}}(\theta)$ on the power $p$ of the three-body regulator is small
 [much smaller than the dependence of $P_{\text{pair}}(r)$ and
$P_{\text{triple}}(R)$ on $p$].
Figure~\ref{Fig12} 
shows the angular distribution function $P_{\text{tot}}(\theta)$
for $N$-boson clusters interacting through 2bZR+3bR6 for various $N$.
The lines from top to bottom at $\theta=0$ are for $N=5, 6, 7, 9,$ and
$13$.
As the number of particles increases, the probability of finding triangles
with small angles decreases but remains finite.
Intuitively, this is because $P_{\text{tot}}(\theta)$ accounts for 
all the trimer configurations and
not just the ``closest trimers''.

Combining the information displayed in Figs.~\ref{Fig6}--\ref{Fig12}, 
the key characteristics of the ground state
of $N$-boson droplets interacting through the
model 2bZR+3bRp with $p \ge 4$
can be summarized as follows:
(i) The dependence of the energy and the structural properties on the
three-body regulator decreases with increasing $p$;
(ii) the dependence of the energy and the structural properties on the
three-body regulator cannot be explained by simple length scale arguments (the
separation of scales is largest for the $p=4$ regulator and smallest for the
$p=8$ regulator);
(iii) 
the pair and triple distribution functions collapse to a
very good approximation to a single curve
if scaled by the binding momentum
of the $N$-body system, suggesting that $1/\kappa_N$ 
and not $1/\kappa_3$ is the
governing length scale for $N>3$.

\section{Results for other interaction models}
\label{secnbodyother}

We now compare the findings for $N$-boson systems interacting
through the model
2bZR+3bRp with $p=4-8$
(see the previous section) with those for $N$-boson systems 
interacting through the models 2bG, 2bLJ, 2b10-6 and 2b8-6.

We start our discussion with the model 2bG, for which the energy
per particle scales, to a very good approximation,
linearly with $N$ for $N \gtrsim 6$ [see diamonds in Fig.~\ref{Fig6}(a)].
The model 2bG has no repulsive core
and is characterized by a single length scale, the width $r_0$.
Using a simple variational Gaussian
product wave function in the single-particle
coordinates, one can readily
show that the ground state energy scales as $N^2$ 
and that the peak density
increases quadratically with $N$.
Indeed, our calculations shown in Figs.~\ref{Fig8}(b) and \ref{Fig10} 
for up to $N=15$ clearly support that the droplet
shrinks with increasing $N$. 
As can be seen in Fig.~\ref{Fig8}(b),
the average interparticle distance quickly decreases to a value
smaller than $r_0$. 
We conclude
that the $N^2$ scaling of the energy 
for the model 2bG
predominantly reflects the absence of a repulsive core
in the potential energy and 
less so Efimov
characteristics.

Next, we discuss the properties of the Hamiltonian
interacting through the van der Waals models 2bLJ, 2b10-6, and 2b8-6.
Our calculations at unitarity 
are performed using the same atomic mass and the same
$c_6$ coefficient for the three models while
the short-range coefficients
are tuned such that the dimer supports a single $s$-wave 
bound state with zero energy. 
For the three-body system, we find
$\kappa_3 L_{\text{vdW}} = 
0.230
$ for the model 2bLJ,
$\kappa_3 L_{\text{vdW}} =
0.233
$ for the model 2b10-6, and
$\kappa_3 L_{\text{vdW}} = 
0.245
$ for the model 2b8-6,
i.e., the three-body binding momentum depends
weakly on the
short-range scale of the two-body potential.
The $N$-body energies per particle, in units of the three-body energy 
per particle, are summarized in Table~\ref{tabdmc}.
\begin{table}[!tbp]
  \centering
  \caption{
    DMC energies for the Hamiltonian with two-body van der Waals interactions
    for $N=4-15$. Columns 2-4 show the scaled energy 
    $E_N/N / (E_3/3)$ for the models 2bLJ, 2b10-6, and 2b8-6, respectively.
    The error bars (not explicitly reported) are around 1\%.
  }
\begin{tabular} 
{c c c c}
    \hline
    \hline
     $N$& 2bLJ & 2b10-6& 2b8-6 \\
    \hline
4 &  3.978 	& 3.953  & 3.960 \\
5 &  7.827 	& 7.841  & 7.887 \\
6 &  11.95 	& 11.99  & 12.12 \\
7 &  16.07 	& 16.15  & 16.40 \\
8 &  20.09 	& 20.24  & 20.59 \\
9 &  23.94 	& 24.15  & 24.69 \\
10&  27.57   & 27.89  & 28.57 \\
11&  31.07   & 31.44  & 32.29 \\
12&  34.37   & 34.81  & 35.86 \\
13&  37.50   & 38.02  & 39.25 \\
14&  40.46   & 41.06  & 42.41 \\
15&  43.27   & 43.97  & 45.46 \\
    \hline
    \hline
  \end{tabular}
  \label{tabdmc}
\end{table}
These energies are obtained by the DMC approach~\cite{lester94}.
Dividing the $N$-body energies by the corresponding
three-body energy, the
energy per particle curves
for the three van der Waals interaction
models nearly collapse [see Fig.~\ref{Fig6}(c)].
This can be interpreted as van der Waals universality 
in the $N$-body sector.
Due to the repulsive core, the energy per particle flattens
around $N =10$, indicating that the system starts to grow
outward, i.e., starts to form a ``second layer'' (of
course, the system
is liquid-like 
and individual layers cannot be identified).
Consistent with this,
Fig.~\ref{Fig8}(c) shows that the average interparticle distance 
first decreases with increasing $N$ and then slowly increases 
for $N \gtrsim 8$.

The dashed line in Fig.~\ref{Fig11}(b) shows the pair distribution
function of the $N=13$ system interacting through the model 2bLJ.
The amplitude in the small $r$ region is suppressed
compared to the other interaction models considered
due to the repulsive two-body
core.
Scaling $r^2 P_{\text{pair}}(r)$ using $\kappa_{13}$ (not shown) does not
bring the pair distribution function for the model 2bLJ in agreement
with the scaled pair distribution functions
shown in the inset of Fig.~\ref{Fig11}(a) for the model 2bZR+3bRp with $p=4-8$.
This reflects the fact that a notably smaller fraction of the 
wave function amplitude resides in the classically forbidden
region for the model 2bLJ than for the model 2bZR+3bRp with
$p=4-8$.

\begin{figure}[!tbp]
\centering
\includegraphics[angle=0,width=0.4\textwidth]{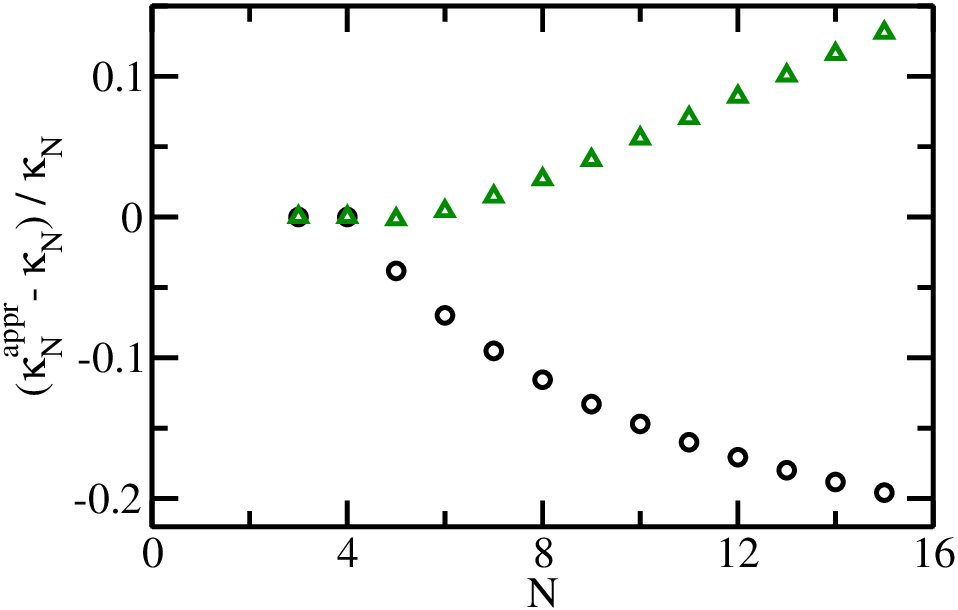}
\caption{(Color online)
  Assessing the applicability of Eq.~(\ref{eq_kappa}) for $N$-boson systems
  with two-body finite-range interactions at unitarity. Circles and triangles
  show the normalized difference 
$(\kappa_N^{appr}-\kappa_N)/\kappa_N$
for the models 2bG and 2bLJ, respectively, as a function of the number of
particles $N$.
 }\label{Fig13}
\end{figure}

As already mentioned in Sec.~\ref{secclusteroverview}, Eq.~(\ref{eq_kappa})
applies, according to Ref.~\cite{kievsky142}, not only to systems with
zero-range interactions but also to systems with finite-range two-body
interactions.
To assess the applicability of Eq.~(\ref{eq_kappa}),
we denote the left hand side of Eq.~(\ref{eq_kappa})
by $\kappa_N^{appr}/\kappa_3$ and
plot the normalized difference between $\kappa_N^{appr}/\kappa_3$ 
and the exact $\kappa_N/\kappa_3$, as determined by our calculations.
Circles and triangles in Fig.~\ref{Fig13}
shows the
quantity
$(\kappa_N^{appr}-\kappa_N)/\kappa_N$ for
the models 2bG and 2bLJ, respectively.
For $N=3$ and $N=4$, the normalized difference is zero by 
construction.
For $N>4$, the normalized difference is negative for the model 2bG and
positive for the model 2bLJ. The deviations from the functional form proposed
by Gattobigio {\em{et al.}} increase roughly linearly with $N$ for the model
2bLJ, reaching 13\% for $N=15$, and non-linearly for the
model 2bG, reaching $-20\%$ for $N=15$. Thus if high accuracy predictions are
sought, then Eq.~(\ref{eq_kappa}) should be used with caution.
\section{Conclusions}

This paper studied weakly-bound Bose droplets at unitarity.
These systems are obtained by adding one atom at a time
to an Efimov trimer or a weakly-bound trimer with
Efimov characteristics.
We carefully analyzed the three-body system and then studied
larger systems.

The three-body ground state of the Hamiltonian with
two-body zero-range interactions and repulsive three-body
potential (model 2bZR+3bRp) is a nearly ideal Efimov state.
The premise was (see also Ref.~\cite{javier10})
that this would allow us to determine the universal
properties of droplets tied to a three-body Efimov state by studying $N$-body
ground states. Somewhat surprisingly, we found dependences of the
ground state cluster properties on the three-body regulator, suggesting
that the ground states become less universal with increasing $N$.
This is a somewhat disappointing finding as the treatment of
$N$-body excited and resonance states, which are expected to exhibit universal
characteristics, is a computationally
much more demanding task. Yet, our study revealed a different
type of universality for these model Hamiltonian. We found that
if the lengths are scaled by the $N$-body binding momentum,
then the dependence on the three-body regulator diminishes
notably. This suggests that the ground states of these systems are 
halo states~\cite{jensen04}, i.e., states whose amplitude is predominantly
located in the classically forbidden region.
The $N$-body binding momentum itself is, however, not---as it would be 
in the case of $N$-body Efimov universality---determined by
the three-body binding momentum, especially not 
as $N$ increases.

Hamiltonian with two-body van der Waals interaction at
unitarity were also investigated. It was found that 
the energy per particle, if scaled by the three-body 
energy, collapses to a very good approximation to a single curve,
suggesting that the short-range details of the van der
Waals interaction impact the three- and higher-body
sectors in a similar manner (i.e., the short-range details are
to a very good approximation ``taken out'' by scaling by the
three-body energy).
The calculations presented were for Lenard-Jones and
modified Lenard-Jones potentials; the latter potentials have
a $-c_6/r^6$ tail but a softer
repulsive core at small distances than typical van der Waals interactions.
We also performed calculations
for (i) the true helium-helium potential scaled by an overall
factor such that the $s$-wave scattering length
is infinitely large and (ii) the true helium-helium potential
with modified short-range potential 
such that the $s$-wave scattering length
is infinitely large (these models were labeled He-He(scale)
and He-He(arctan) in Ref.~\cite{blume15}).
The energy per particle curves for these systems, which have a more
complicated long-range tail, also collapse, to a very good approximation,
to the same curves as those for 2bLJ, 2b10-6, and 2b8-6
if scaled by the three-body energy.
The structural properties, specifically the pair and triple
distribution functions, for the van der Waals systems do
not collapse to the same curves as those
for the 2bZR+3bRp model with
$p=4-8$ if scaled using the $N$-body binding momentum $\kappa_N$,
suggesting that a good portion of the
wave function amplitude of the van der Waals systems is located in
the classically allowed region.

In the future, it would be interesting to extend the
calculations presented here to excited and resonance states.
We expect that the $N$-body properties become
universal if sufficiently high excitations are being considered.
In the four-body sector, e.g., Deltuva~\cite{deltuva10}
extracted the universal numbers for $\kappa_4/\kappa_3$ 
by going to high-lying resonance states
(in this case, ``high-lying'' means third or higher resonance states). 
Extending calculations such as those conducted by Deltuva to
$N>4$ is, however, challenging. It would also be interesting to extend
the studies presented in this paper
to finite $s$-wave scattering lengths and to Bose droplets with
an impurity.

\label{secconclusion}
{\em{Acknowledgement:}}
We thank Aksel Jensen for suggesting that we think about 
weakly-interacting systems in the context of classically 
allowed and classically forbidden regions.
Support by the National
Science Foundation (NSF) through Grant No.
PHY-1415112
is gratefully acknowledged.
This work used the Extreme Science and Engineering
Discovery Environment (XSEDE), which is supported by
NSF Grant No. OCI-1053575, and the
WSU HPC.
\bibliographystyle{apsrev4-1}
\nocite{apsrev41Control}
\end{document}